# “Sora is Incredible and Scary”: Emerging Governance Challenges of Text-to-Video Generative AI Models

**Kyrie Zhixuan Zhou** University of Illinois at Urbana-Champaign | zz78@illinois.edu
**Abhinav Choudhry** University of Illinois at Urbana-Champaign | ac62@illinois.edu
**Ece Gumusel** Indiana University Bloomington | egumusel@iu.edu
**Madelyn Rose Sanfilippo** University of Illinois at Urbana-Champaign | madelyns@illinois.edu

## ABSTRACT
Text-to-video generative AI models such as Sora OpenAI have the potential to disrupt multiple industries. In this paper, we report a qualitative social media analysis aiming to uncover people’s perceived impact of and concerns about Sora’s integration. We collected and analyzed comments (N=292) under popular posts about Sora-generated videos, comparison between Sora videos and Midjourney images, and artists’ complaints about copyright infringement by Generative AI. We found that people were most concerned about Sora’s impact on content creation-related industries. Emerging governance challenges included the for-profit nature of OpenAI, the blurred boundaries between real and fake content, human autonomy, data privacy, copyright issues, and environmental impact. Potential regulatory solutions proposed by people included law-enforced labeling of AI content and AI literacy education for the public. Based on the findings, we discuss the importance of gauging people’s tech perceptions early and propose policy recommendations to regulate Sora before its public release.

## INTRODUCTION
As artificial intelligence (AI) continues to permeate various aspects of our daily lives, integrating advanced AI systems raises significant questions and prompts diverse reactions within society. In particular, the remarkable advancements of Generative AI (GenAI) have transformed the domain of Human-Computer Interaction (HCI) (Muller et al., 2022). Sora OpenAI (Sora) is a recently announced advanced GenAI tool capable of crafting realistic videos based on textual input (Sun et al., 2024; Wang et al., 2024), which has sparked considerable interest and speculation regarding its potential impact on societal dynamics, governance structures, and social norms in society. Building upon advanced natural language processing (NLP) and machine learning (ML) techniques, Sora is equipped with superb generative abilities and holds the potential to revolutionize industries (Guan et al., 2024), reshape social interactions, and redefine the boundaries of human-AI collaboration (Waisberg et al., 2024).

As society stands at the precipice of integrating Sora into various domains, such as healthcare, education, and art, the need to comprehend the intricate dynamics of public perceptions and the broader societal implications becomes increasingly urgent. Effectively navigating the complex intersection of technology and society requires a deep understanding of the ethical, social, and cultural dimensions that accompany AI integration. Addressing the initial reactions and concerns may mitigate the potential risks and safeguard the values and interests of individuals and communities after Sora’s release. Mogavi et al. (2024) qualitatively analyzed Reddit discussion around Sora to understand people’s perceived applications of and concerns about Sora’s integration prior to its public release. We collected Sora-related discussions from a different social media platform, X (formerly Twitter). In addition to discussions about Sora-generated videos, we also included discussions related to a comparison between Sora videos and Midjourney images, and discussions on ethics initiated by an artist whose work was copied by generative AI, in our analysis. We additionally focused on governance solutions to people’s expressed concerns.

We aspired to answer the following research questions:

**RQ1.** What are the primary reactions, opinions, and concerns people express regarding Sora’s integration?

**RQ2.** What are the perceived implications of Sora for governance structures and social norms?

Our qualitative analysis revealed people’s concerns about Sora’s integration into content creation-related industries, including OpenAI’s for-profit nature which makes their claimed safety measures less credible, the blurred boundaries between real and fake content, degraded human autonomy, data privacy, copyright issues, and environmental impact. Optimism was not completely missing, in that generative AI (e.g., Sora and Midjourney) seemed to converge to similar outputs based on existing training data without exhibiting creativity. People further proposed solutions to mitigate potential negative impacts brought by Sora, such as AI literacy education for the public, and law-enforced labeling of AI-generated content. Based on the findings, we proposed policy recommendations to some of the uttered concerns, namely, AI watermarking, leveraging existing laws, enhancing accessibility of Sora, and adopting an entrepreneurial approach to regulation.

We contribute to the emerging literature on Sora and text-to-video GenAI in two ways. First, we explored people's reactions to and concerns about Sora's integration, and their potential solutions to prevent its negative impact. Second, drawing on social media discussions, we proposed policy recommendations to harness Sora's negative impact on human jobs and unlawful/unethical domains.

## RELATED WORK

### The Impact of Sora: Benefits, Challenges, and Public Perceptions

Sora is an AI tool for text-to-video generation by OpenAI, whose demos show surprisingly good quality (In, 2024). Sora has blurred the lines between tangible and virtual realms and even "prompted viewers to ponder their very existence" (Cavalcante, 2024, p.2). Li et al. (2024b) created a benchmark to assess the quality of AI-generated videos based on their adherence to real-world physics principles and showed Sora's advantage over other text-to-image generation tools. In recent literature, scholars have predominantly focused on addressing the challenges, benefits, and public perceptions of Sora.

#### *Benefits*

Sora is poised to disrupt multiple industries, including movies, education, gaming, healthcare, and robotics, with potential benefits for individuals (Liu et al., 2024). According to Lin (2024), Sora may serve to broaden access to mental healthcare services. In the realm of robotics, Sora's capability to simulate intricate world dynamics proves crucial for advancements in autonomous driving (Guan et al., 2024), such as scenario generation (Li et al., 2024a). Democratization of art creation is a notable affordance of Sora – "Everyone will be a filmmaker" (Cowen, 2024), which holds promise for broadening participation in artistic endeavors. Researchers speculate that medical professionals can use Sora to create informative videos for patients and the public (Waisberg et al., 2024). Sora can also become a potentially powerful tool for education and libraries, offering opportunities for diverse learning modalities, creativity, and critical thinking, while at the same time raising challenges such as bias mitigation and equitable access (Adetayo et al., 2024).

#### *Challenges*

Sora may introduce ethical quandaries concerning privacy and security, copyright, disinformation, and truth integrity. AI's enhanced video generation capabilities may ease the creation of adversarial attacks (Hannon et al., 2024). Copyright infringement is a paramount concern brought by Sora – it might have leveraged copyrighted materials such as videos for training (Karaarslan & Aydın, 2024). Other concerns raised by the researchers included misuse of AI-generated content (e.g., deep fake videos) and environmental costs. Rodríguez (2024) discussed the likely "ultimate solipsism" introduced by the rise of "self-movies," where protagonists are the creators themselves, leading to implications of quality degradation. Moreover, the duality of Sora was expressed by Cavalcante (2024), who argues that Sora simultaneously indicates democratizing creation, i.e., anyone can become a visual storyteller regardless of their technical or financial status, and ethical dilemmas such as disinformation that put the integrity of truth at risk.

#### *Public Perceptions*

Some researchers have started investigating the public's perceptions of Sora even before its release (Mogavi et al., 2024): both people's envisioned applications of Sora and expressed concerns about this emerging technology were captured by qualitatively analyzing Reddit discussions. People predicted that Sora would democratize video marketing and gaming content production, enhance educational content by generating illustrative videos, and substantiate people's creativity and storytelling abilities. In the meantime, Sora threatens creative and artistic jobs and perpetuates social biases from the dataset it is trained on. We extend this work by examining discussions on a different online platform, comparing discussions around Sora and Midjourney, and focusing on governance strategies to regulate emerging challenges.

### Tech Hypes

A way to look at how a technology such as Sora would progress is through the lens of technology adoption models including the Theory of Diffusion of Innovations, the Theory of Reasonable Action, Decomposed Theory of Planned Behavior, and the Technology Acceptance Models 1, 2, & 3 (Lai, 2017). These models look more at perception and the adoption of the product through the consumer lens but do not examine expectation and attention. The perceptions of the intrinsic characteristics of the technology produce an S-shaped curve of adoption (Shi & Herniman, 2023) characterized by five adopter categories: innovators, early adopters, early majority, late majority, and laggards, after which the point of diminishing returns of adoption is reached (Rogers, 1962). A pattern of interest is that newer technologies appear to be adopted faster than older technologies (Saxonhouse et al., 2010).

Prior to the release and widespread use of a product, and its early adoption, its mention in both media and technical texts helps it get attention and investments though such attention is a double-edged sword because it also leads to buildup of expectations to an extent beyond what is possible to deliver, leading to damaged credibility and reputation (Brown, 2003). The word "hype" has been used to represent this cycle, and this most famously includes the Gartner tech hype cycle methodology developed by Gartner Inc., a technology research and consulting firm, in 1995; Gartner uses the hype cycle in its consulting reports to understand the evolution of the hype around emerging technologies. Gartner's hype cycle has five stages: Innovation Trigger, Peak of Inflated Expectations, Trough of Disillusionment, Slope of Enlightenment, and Plateau of Productivity. At the innovation trigger stage, the technology is at the proof-of-concept while in the second stage, there are very few practical implementations despite expectations peaking. Thereafter, the third trough stage is a sharp decline of interest and many business failures. It is only after this that the technology picks up again more maturely and finally reaches mainstream adoption in the final stage (Gartner Inc., n.d.). We can infer that early expectations being inflated beyond what could be reasonably delivered is the expected norm of evolution of new technologies. It is further understood that the time needed for the full hype cycle might vary and while "normal technologies might take five to eight years, fast-track technologies might only take two to four years for maturity" (Dedehayir & Steinert, 2016). The bell-shaped hype curve produced is due to human irrationality, mainly the following: attraction to novelty and love for sharing, social contagion, and heuristic attitude in decision-making (Fenn & Raskino, 2008). This rapid boom and bust contrasts with the actual maturity of a technology, and technological growth is a combination of continuous or incremental changes, and discontinuous or radical changes, together resulting in a predictable technology trajectory even though the changes themselves might not be predictable in advance (Dosi, 1982).

Sora could be looked at as wholly a new technology, but it could also be regarded as a continuation of generative AI models that have taken the world by storm in the past couple of years. In that regard, it is either going to have its own hype cycle and will go through a sharp peak and then crashing expectations and disappointment, or it could itself be seen as an innovation in the larger hype cycle of GenAI. A particular autoethnographic investigation on GenAI by King & Prasetyo (2023) demonstrated that such a cycle could also manifest itself at the individual level: it can be inferred that the larger societal cycles are an agglomeration of such individual cycles. It is also possible that there may be multiple peaks and troughs of visibility and expectations rather than a single one and this may not conform to a hype cycle at all (Dedehayir & Steinert, 2016; Steinert & Leifer, 2010). In this study, we aim to capture people's initial reactions to and concerns about Sora qualitatively.

## METHODOLOGY

### Data Collection

To address our research questions, we manually gathered substantial comments under four popular posts relevant to public sentiment or governance of Sora on X (formerly Twitter). The first and second posts contained Sora videos released by OpenAI – the first post was by a tech influencer, while the second post was by OpenAI itself. The third post compared Sora videos and Midjourney pictures created with the same prompts, emphasizing visual similarity. The fourth post was by an artist complaining about Midjourney's copying behavior. These posts were selected for analysis based on their high levels of engagement, indicated by the large numbers of views and comments discussing Sora or public reactions to GenAI. By including discussion about and comparisons with Midjourney, a text-to-image AI tool, we were able to uncover how text-to-video models like Sora presented new challenges to AI regulation and society compared to text-to-image models. Descriptives of the posts are presented in Table 1. The comments were put in a hierarchical structure to reflect interaction dynamics. Our dataset includes 93, 139, 39, and 21 comments respectively from these posts. We were able to arrive at this truncated list from the original number of comments by excluding comments that included little information. For example, although there were nearly 10 thousand comments under Post 2 by OpenAI by early March when we collected the data, a large majority of the comments were short and not informative (e.g., "this is great"), making our data collection process more feasible.

### Data Analysis

We employed qualitative thematic analysis (Braun & Clarke, 2012) to analyze textual data. The thematic analysis involved identifying recurring themes, patterns, and categories within the comments, enabling us to gain insights into the predominant discussion topics and the underlying sentiments expressed by users. Through qualitative analysis, we aimed to uncover the diversity of opinions, concerns, and perspectives expressed by users regarding Sora (RQ1) and its implications for governance and society (RQ2). By immersing ourselves in the data, we captured the richness and complexity of the discussions, gaining a deeper understanding of the factors influencing public sentiment towards AI technologies. Below, we use anonymized quotes to illustrate our findings. We retained grammar errors and typos in the quotes but partially censored swear words with [***] and removed emojis.

| Number | Poster | Topic | Post time | # Views | # Comments | # Retweets | # Likes | # Bookmarks |
|---|---|---|---|---|---|---|---|---|
| 1 | Tech influencer | Sora videos | Feb 15, 2024 | 9.4M | 437 | 3.2K | 19K | 8K |
| 2 | OpenAI | Sora videos | Feb 15, 2024 | 94.3M | 9.9K | 76K | 142K | 40K |
| 3 | Tech influencer | Sora videos vs. Midjourney images | Feb 16, 2024 | 11.2M | 674 | 3.9K | 31K | 11K |
| 4 | Artist | Midjourney copying | Mar 8, 2024 | 2.7M | 127 | 9.2K | 63K | 3.2K |

**Table 1.** Metadata about four X posts

## FINDINGS

### Varied Assessment of Video Generation Quality

People had varied assessments about the video quality Sora afforded. While some regarded Sora's video generation capability as revolutionary, others pointed out failures in simulating the physical world and its lack of creativity.

#### *Revolutionary*

Many people are amazed by Sora's video generation quality: *"These videos looks so real.", "I cannot believe we are here already."* The videos are also perceived as of higher quality than videos generated by other AI: *"I have been watching many AI videos recently but this is just next level. That small figurine video inside a glass bowl is something we see from an old camera. Can't wait to test it out!"* Some people particularly praised the physical features captured in videos, *"the physics etc are just too good."*

Sora was seen as a revolutionary technology, *"It has come to a point where it is hard to imagine what AI will be able to do next."* AI-generated videos may make the virtual world cooler for people, *"W[***]. Went to bed and woke up to fake puppies playing in the snow. This is mindblowing. We're reaching the point where the virtual world will be much cooler than real life for 90% of people."* Reality and simulation were blurred, *"AI like Sora is pushing the envelope, turning text into lifelike scenes and inching us closer to a world where reality and simulation blur. It's a glimpse into a future where the simulation hypothesis might not just be a theory. Could we already be living in a simulation? Sora's advancements hint at the possibility."* Some suggested AI may have had access to alternative realities, *"What if the Ai has already access to alternative realities and is capturing those footages out there and that's the reason it looks so realistic and we cannot see the difference."* These statements are reminiscent of science fiction and revealed some people's perceptions of the advanced nature of GenAI.

#### *Just So-so*

Some people were skeptical and thought OpenAI might have used videos that matched the prompt, *"How do we know its not just giving out unaltered training data that matches the prompt? a dog in the snow video is not uncommon."* One person pointed out that many of the positive comments may be bot-generated, *"Just a load of bot accounts saying how amazing it is and that they can't wait to try it. It's s[***]."*

To some, the videos were of "game engine-level quality", *"I don' think it even considers physics in that sense, like if it were a game engine."* Failures in capturing physics were pointed out, e.g., *"Dog skins and snow doesn't follow physics correctly."* Many people found that the walking woman's legs switched sides several times in a video.

Some people thought the Sora videos were merely a product of the huge computational power or datasets, *"Can you guys understand that these models all are a patchwork of REAL,PRE-EXISTING CONTENT? There's nothing strange or impressive, it's just how these algorithms work (INFERENCE/ENCODERs/DECODERs). What's impressive is the computational power at disposal. But I can't wait for people to start digging into these DATASETS to DEBUNK the "magic" once and for all... Come on people, they're SELLING a PRODUCT... Don't fall for it."*

When comparing Sora and Midjourney, people found they converged to similar output, *"This just shows that most foundational models are going to converge to a similar output. This should give some AI hype pause."* Proprietary data was seen as a key to training high-performance models, whereas all models currently are trained on similar public data, *"It's now down to who has the most robust proprietary data. The issue now is all these models are* *trained off the same data sets."* With all the AI-generated content online as potential training data, the convergence would only get worse, *"Especially as the internet becomes filled with "noise" in the form of AI outputs, convergence seems inevitable."*

After noticing similar generated content by Sora and Midjourney with the same prompt, people also believed that GenAI was not creative and would not replace human workers, *"I know folks are worried, but AI is going to just fool some CEOs into laying off in the short term till they realize no one will buy what is essentially uninspired clip art. Going to suck until the CEOs rehire folks though."* However, another person pointed out the similar convergence in human creativity, *"Humans don't converge to similar ideas, behaviours....?"* In addition, people doubted GenAI's role in stimulating people's imagination, *"Since this 1 is the ONLY moderately creative 1 I've seen so far, I wonder if this tech is going to stimulate people's imagination or atrophy it, going to be used mostly to falsify mundane & ugly s[***], instead of coming up with imagination defying beauty?"*

### Impact on Industries and Professions

Sora has the potential to disrupt movie and content creation industries, as expressed by many. Ambivalence was demonstrated by this quote, *"Sora is incredible and scary."*

People agreed that the movie industry and Hollywood would be directly impacted by the prevalence of video-generation AI like Sora, *"Hollywood is about to implode and go thermonuclear"*, *"RIP the entertainment industry. In a few months, people will make "Avatar" on their phones."* One person even imagined the collaboration between Sora and Midjourney, *"Midjourney can generate the thumbnails while Sora the movie"* Some started to imagine a new form of streaming industry, *"Hollywood's days are numbered! OpenAI will make everyone an artist. From now on, anyone will be able to create and launch their own movies on the market. I feel that a new type of global streaming service will emerge. I hope Netflix is paying attention…"* Sora may democratize movie making, which has unpredictable implications. One person raved, *"Thank you, OpenAI has literally democratized cinema. Now, everyone can be an artist and create their own movies going forward. This will drastically reduce the exorbitant expenses of Hollywood, especially for low-quality content with high costs.*" However, the democratization of movie creation may also lead to low-quality content, *"Democratize: Discard all talent through stealing people's work then allow anyone who don't know a thing about the art to create their own slop. Wow I love this new democratized cinema world. xD Have fun swimming in your own c[***]."*

This impact may extend to the rest of the media production industry, content creators, and artists. On the bright side, Sora may enable writers to do new forms of storytelling, *"Perhaps one day regular people, as authors, will be able to create feature-length films from (only) their writing. It would be a refreshing restart to finally wrestle the art form away from Hollywood and allow real stories to be told again."* Sora may democratize content creation for average people, *"Awesome tool. Tools like this integrated with no-code will truly democratize what regular people can create and push out online. This stuff is amazing and that exponential growth this year will be insane."* People liked the fact that they could create whatever they wanted, *"Great! Now I can create the video I've always wanted to make, where a ton of really famous celebrities confirm that software quality goes waaay beyond "correctness" to requirements, into the realm of the emotive experience of using it and the value it brings. Oh and ethics.*" Sora may also nudge artists to pursue aesthetics to stay on par with GenAI, *"This is beautiful and it will inspire artists to make old school movies to contrast with this aesthetic."*

The other side of this coin, however, is that it implicates job loss for artists, *"I don't think y'all realize how many artists you're f[***] over right now"*; the rise of AI content farms, *"I would like to miss it because it threatens every single creative profession. Music creation, art, story writing, screenwriting, photography, and film. Unless you are at the highest production level where protections exist, it'll be impossible for human creators to get traction."*; and the frustration of art students, *"AI is making me feel i'm wasting my time studying arts. At this point I fear my future is gonna be affected by this medium, the fact that this fear is getting stronger with each new i see about AI 'progressing' sickens me, and I'm sure I'm not the only one."* Some people pointed out an unfair game between GenAI and human artists, i.e., GenAI has been stealing artists' work as its training data, *"It hasnt democratised anything. It has stolen peoples data without consent to profit of it. Everyone could already be an artist some are just to lazy too."* When an artist shared her experience of being copied by Midjourney, another artist was frustrated and questioned why AI evangelists could not see this dark side of GenAI, *"Breaks my heart. I'm an amateur artist. I know that like most things it requires practice and patience, I chose to make my career elsewhere, but I'm sure I'd be pissed if other people bastardised my professional efforts. I simply can't see why the AI evangelicals don't see it."* One person vividly described these implications as communism in the art field, *"Welcome to communism. No longer is the ability to create art held by the elite, your skills have been equitably redistributed to all the workers comrade. Now we are equal."*

Some people are optimistic even with all the negative impacts, thinking Sora would not replace artists and content creators in the short term since it was not creative, *"It's never replacing real artists. Machines only do loops so that means that they only repeat what has been made. Only Humans are truly creative"*; did not have feelings, *"True art connects with feelings. So far at least, neither machines nor machine code contain feelings. There is something truly algorithmic in the ai generated at I have seen so far. Do love the drone footage of the gold rush though. My opinion"*; and did not recreate the human touch, *"Even as machinery comes to the scene, there's always gonna be a demand for man-made stuff. It can never truly recreate that human touch so integral to the craft. Keep doing what you're doing, keep pursuing what you want to do"*.

### Emerging Governance Challenges

People commonly shared negative sentiments regarding OpenAI as a troublemaker, *"DeepMind is solving problems, OpenAI is creating them, lol"*; numerous AI bro posts, *"These AI bro posts are exhausting."*; unpreparedness of society in the face of governance challenges, *"We aren't ready for this as a society"*; and negative outcomes outweighing benefits, *"This will be used in the worst possible way. I know it's out there and impossible to stop. The negative outcome of this, outweighs by far the benefits to society. Dangerous times ahead"*.

Some people are less worried about these governance challenges brought by emerging technologies, comparing Sora to cars and movies, *"... just like 135 years ago when people got scared of the first moving pictures ..." "Yeah sure, just like around 1910 when cars started to replace horses and farriers lost their jobs, or when Netflix killed Blockbuster. It's happening, get over it."* Some believed morality could always catch up with technological evolution, *"Which has almost always been true with science and technology - the science always comes way before the ethics and and, sometimes, morality can catch up."*

Governance challenges discussed included OpenAI's for-profit nature, the authenticity of content, taking people's jobs, training data, power consumption, and the copyright of AI-generated content.

#### *For-Profit OpenAI*

People pointed out the for-profit nature of OpenAI, which makes generative AI such as Sora close-sourced, *"What shocks me is not just the Quality but the level of Secrecy that OpenAI was able to maintain around this project is insanely impressive."* Others thought OpenAI did not aim to research AI for the public good but only pursued profits, *"Ah yes for a company supposedly dedicated to researching AI and bringing it to the public I sure love the level of secrecy involved at the company, and it's not at all terrifying that a company with this level of secrecy wants 7 trillion dollars for further AI development."* People thought OpenAI did not care much about ethical concerns arising from Sora, *"They just don't care. And they don't care about the slew of fake-videos and propaganda and computer generated porn of real people and all of the people they'll hurt that way. They don't care about anything as long as they get to line their pockets."*

OpenAI claimed to ensure safety before releasing Sora, *"We'll be taking several important safety steps ahead of making Sora available in OpenAI's products. We are working with red teamers — domain experts in areas like misinformation, hateful content, and bias — who are adversarially testing the model."* However, some people questioned their use of artworks for training, *"As part of this, how do you take into account that AI is derivative of other people's existing photography, art, and concepts? What are the ethic and protections around that governing your business?"*; the definition of safety risks mentioned by OpenAI, *"How do you determine what is considered 'misinformation, hateful content, and bias'? There is no such thing as 'hateful content' 'Misinformation' is subjective 'Bias' is subjective Explain yourselves @OpenAI"*; the distribution of responsibility for ensuring safety, *"What kind of person is a domain expert in misinformation, hateful content and bias? Did you create a GPT of Adolf Hitler and Joseph Goebbels?"*; and the creation of Sora itself, *"We don't want it. Period. Nobody asked, and it's a danger for our children, artists, & economy as a whole. I suggest you shut this down before the lawsuits come. Because they will."*

#### *What is Real? Falsified Evidence, Pornography, and Propaganda*

Some people claimed that they would no longer *"trust for videos on the web"* since *"nothing is real with AI."* It was hard to tell AI- vs human-generated videos, *"Now people are releasing real videos saying it's AI, no one knows the difference."* However, others argued that things were not real even before AI, *"You don't know what's real now.... If you're taking anything you see in the main stream media at face value, then idk what to tell you. 99% of it is spin, bias, or even flat out wrong. AI used for nefarious purposes will just be an extension of what's already happening"*. People also expressed that humans have been bad at distinguishing between fake and real all the time, *"I mean, most people fall for the worst faked videos all the time...so, nothing's really changed."* They thought technologies were not to blame, *"Lighters allow anyone to set a fire anywhere. This dangerous tool should be strictly controlled, otherwise we will all die in fires. how right it sounds."* They also believed technological solutions would come to the rescue, *"A guy said years ago on some podcast i watched, as good as technology is at making something fake there's as good technology identifying if somethings fake."*

Practically, some people raised concern about falsified videos as court evidence, *"this is all fun and games until you end up in court watching a 60 second video evidence of yourself committing a crime you've never done."* Another person replied with actional steps to falsify such a video, *"Now you can Google a random person's name, get just a profile pic and in a few short sentences create a video of that person doing anything you want. You only need a starting photo, rather than a full video of someone actually committing the crime to add their face to."* Technologies such as blockchain may help such videos seem non-AI and more admissible, *"Someone will make security cameras that are integrated with blockchain technology and embed authentication directly into the videos so they can be deemed unaltered/non-AI and still admissible."* Some fear that innocent people may be put into jail, *"This tech will put people in prison with falsified evidence. Not there yet, but not far off at all."*

Video generation will also facilitate easier creation of porn, *"Porn is about to get crazy!"*, *"Porn industry about to go spiral into alternate universes of weird."* With simple prompts, video generation AI such as Sora will be able to generate pornography, *"I'm obviously missing something. Please explain to me in simple words why all this 'hyperrealistic' imaging is interesting for anyone except criminals and pornographers. Do you really think 'a fat girl from Ohio with her father's videocamera' (Coppola) will create a true masterpiece?"* Existing platforms for sex workers to produce pornography may be impacted, *"The days of OnlyFans will be over soon then"* Implications for child pornography are concerning, *"you are facilitating child pornography and you know that (: my question is how you people are able to sleep at night knowing what you're doing."*

Video generation AI may further interfere with elections with propaganda videos, *"Fabricated political propaganda will be all over the internet."*, *"Looks like the 2024 election is going to be fun."* Another person disagreed and argued that AI is not the one to blame but the underlying vulnerable democratic system, *"If generated 10 second videos endanger democracy even though it's public knowledge that such video generation is possible, then don't you think the system was brittle to begin? I don't think generative AI is to blame for that."*

### *Replacing Humans and Small Businesses*

People complained that instead of benefiting humans, AI is often used to replace poor people, *"AI is a good representation of how humanity fails to use innovation to progress. We COULD automate menial jobs, implement UBI, and give humans the freedom to create and invent, but instead we replace The Poors™ with barely functioning machines and then leave human to starve."* AI kept squeezing people into smaller boxes of job options, *"Each new automation technology squeezes humanity into a smaller box; and then you see new standard of 'normal' people complain about increases in clinical diagnoses for types of people who no longer have a modern purpose."* Some thought technologies evolved so fast that we had not thought thoroughly about their implications or set rules, *"Can we wait until there's some more rules in place before developing stuff like this please? As of right now, this could damage so many people's livelihoods."*

There was a heated debate between human autonomy and technological evolution. On the one hand, *"You're literally hurting jobs with this."* On the other hand, *"OK let's stop technological evolution because...jobs"* Optimists believed creative jobs or jobs requiring creativity and human touch could not be easily taken by AI, *"Analytical/automated/repetitive tasks are likely to be replaced regardless. Things AI can't do: imagination, creativity, drawing on life experiences, intuition, emotional connection, manual dexterity, etc. Those jobs will receive more boost in the future."*, *"AI might take over many human tasks, but it falls short when it comes to building a community. The human touch in marketing is irreplaceable, and AI will never match it."* According to them, AI could also free people from monotonous jobs, *"AI could be used to reduce the human necessity to work monotonous jobs and allow us to explore our creativity."*

Small Businesses and Startups without comparable computational power will be killed by tech giants such as OpenAI, *"OpenAI just can't stop killing startups."*, *"bro just stop you killing everyone'e business" "Sam Altman has more founders blood on his hand than any other CEO in history of mankind …" "how are we supposed to keep working on our little SAAS product after this? I'd be less anxious even if aliens landed on Earth."*

### *Copyright of Input and Output Data*

Training data was a major concern for many. Some feared that YouTube or TikTok videos have been used for training, which has been written into the platforms' terms of service, *"My conspiracy is I feel TikTok, etc has been built to feed these ai tools and there's language in the terms that allows them to own certain rights to videos uploaded."* Overall, people thought OpenAI was not transparent about where they obtained the training data and suspected they had scraped images and videos from around the internet without consent or *"used copyrighted material without paying for it."* People may unintentionally give up their data rights by not reading the terms of service closely. When one person asked, *"who approved their content to be scraped and put into the training?"* Others replied, *"You did. When you accepted the terms of service.", "Read the entire terms of service."*

The similar output by Midjourney and Sora indicates the importance of proprietary training data like YouTube or X, *"All things being equal, all models eventually get to parity without proprietary training data sets."* Some thought Google had an advantage for the massive amount of data it owns, *"Idk how the legal stuff would work but google is sitting on a f[***] goldmine."* However, others pointed out that training on social media data may lead to bias in AI, as it learned human bias, *"Using social media to train your model on is very dangerous. It has been tried in the past and the ai always was trained by users to become racist within hours."*

Training on artists' and content creators' work, or normal users' content, was seen as stealing, "*Stop stealing people's work.", "So which animators art did you steal for this?" "Prompt: Steal the work of artists who put their lives into the work, then create a system where none of them can work again and feed their families, also do so without warning, so that only we can profit."* Others held different views, believing learning is not stealing, *"Every artist cobbles together other people's work subconsciously from every image they ever saw and has since the stone age. The AI model uses what it learned from pattern recognition which ties key words / sentences to visual concepts / patterns. How is it different?"* However, as some pointed out, such views have *"confused the usage of material that requires a license with the idea of human learning."*

People were also wondering who would own the copyright of AI-generated content. Some thought future copyright law would favor AI developers over AI users, *"In time, law will benefit AI program developers. They will get the copyright, not the 'ai artists'. Mark my words."* Others did not mind the fact that they may not be able to copyright AI-generated content since the sunk costs were low with minimum human efforts in the generation process, *"Does it matter when you can generate new content again and again without the huge sunk costs that the traditional content industries have?"*

#### *Power Consumption*
There was a tension between Sora's performance and power consumption, *"What is the purpose of that tool? Generating videos is awesome, but do we need that amount of power consumption to do that?"* This has implications for global warming, *"how many carbon emissions were released to produce the electricity that powered the servers that rendered this AI generated video??? tl;dr: how much did you contribute to global warming just to produce this c[***]?"*

### Potential Solutions
Most people were in favor of strict regulation and AI literacy education to hold Sora accountable and limit its negative impact.

#### *Law-enforced Labeling of AI Content*
Law-enforced labeling of AI-generated content and tools for identifying AI-generated content are deemed important approaches to regulation, *"Laws are needed immediately. Any AI-modified video or image should be required to include a disclaimer symbol, mark, or tag of some kind that makes it obvious—even in films, artwork, and written content. Extreme penalties are needed for any who break these rules. Everything released should have a clear source or otherwise be unshareable. If we aren't extremely careful, we soon won't be able to believe our own eyes."* Labeling of AI-generated content can facilitate copyright protection in the meantime.

However, many people found law-enforced labeling of AI-generated infeasible for adversarial manipulation by users, *"People would just remove the watermarks or crop the videos"*; distrust in the government who only acted on behalf of rich people and companies, *"They haven't responded because they don't care. Governments don't have our best interests in mind and they definitely have no regard for creators. They make new laws for corporations only"*; politicians' poor tech literacy, *"It doesn't help that a huge percentage of our most powerful politicians were too befuddled by new technologies to figure out how to program their VCRs 20 years ago"*; enforceability of laws, *"Agreed, but like most regulations requiring watermarks/notations/etc. I worry that, while it may discourage the majority of problems, laws won't be enough to completely combat this type of conduct"*; and the slow nature of legislation update, *"Even if they do care, laws always take ages to create and implement in any capacity on purpose. Unfortunately it's far too slow for how fast things are moving in the world of tech and stuff."*

A minority opinion was that guaranteeing safety through strict law indicated *"taking out all the fun"* and *"restricting freedom of speech."* They believed such guardrails might drive Sora unusable and instead, only illegal content should be limited, *"Many of the limitations you have put in are making the product unusable for the majority of users. The only content that should be limited is illegal content as what is 'misinformation', 'hateful content', and 'bias' is different based on human interpretation, time, and other factors."*

*Educating the Public*

In addition to law enforcement, some argued education was a more sustainable approach to mitigating negative impacts of emerging technologies, *“No. We don’t need more laws. We need citizens who can think and reason. The problem of ‘fake news’ (or spin) has been with us for centuries.”* Education is important because AI technologies will be used to nudge people to do bad things, *“We need to educate the public on AI and make it a priority because this technology is going to be used to manipulate and socially engineer people into doing very bad things.”* Source verification is one of the important topics for education, *“Needs to be verified by reliable source (?) or a live happening with large audience present.”*

## DISCUSSION

Mogavi et al. (2024) utilized social media (Reddit) data and a qualitative analysis to understand social media perspectives regarding envisioned applications of Sora and people’s concerns about its integration. Envisioned applications included video marketing, gaming asset creation and cinematics, educational content creation, and digital storytelling. People’s concerns included threat to creative jobs, bias, harm to art, and unpredictability in creative workflows. Compared to Mogavi et al. (2024), people's envisioned application and perceived impact of Sora in our study focused on movies and content creation in general. People expressed a wider range of concerns, including (1) OpenAI’s for-profit nature, which made their claimed safety measures less credible, (2) the blurred boundary between real and fake content, which had practical implications such as falsified evidence in court, pornography, and political propaganda during elections, (3) replacing human workers and small businesses, (4) copyright of input and output data, and (5) power consumption and environmental impact. We further gauged people’s perceptions of potential regulation and solutions to these challenges, namely, law-enforced labeling of AI-generated content, and educating the public. We generated a word cloud based on our collected data using NVivo R.14.23.0, highlighting discussions on art/artists, laws, and content, as shown in Figure 1, with larger font sizes indicating more mentions in the dataset.

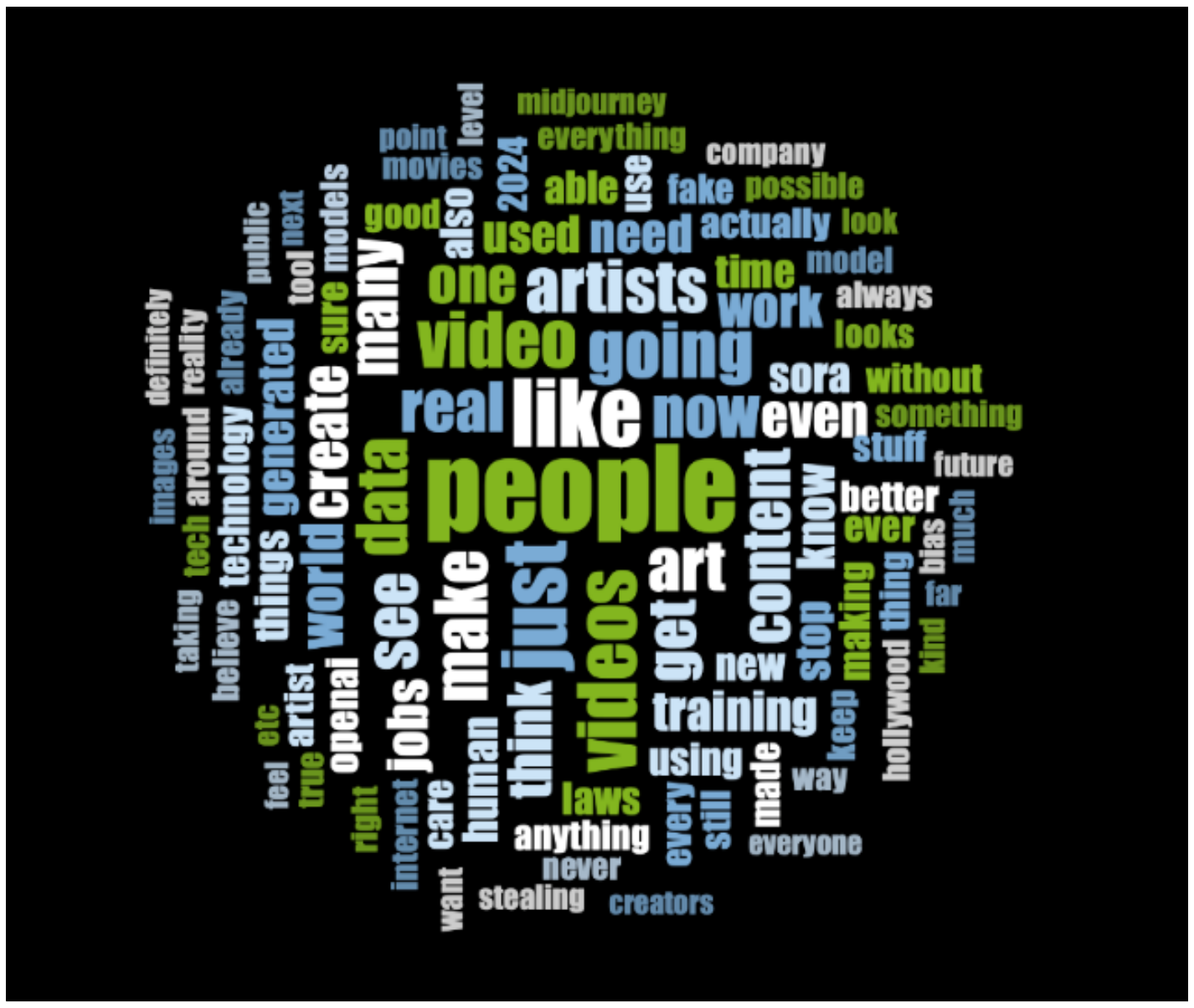

**Figure 1.** Word cloud based on total mentions

## Toward Gauging Tech Hype around Sora

Our work on the public discourse around Sora is a suitable representation of tech hype at the inception stage. With Sora currently at the innovation trigger stage, as the actual product has not been publicly released and only a teaser has been provided by OpenAI, people were nevertheless excited about this proof-of-concept technology, evidenced by heated discussions on Sora-related topics on social media. They had interestingly formed perceptions and expectations on the basis of the very limited information they have about Sora by drawing on past experiences with AI. We find that there is a directionality to the expectations around Sora, with most comments taking a clear positive or negative position, and expectations and perceptions were also logical/ fact-based or emotional, resonating with prior research on tech hype expectations (Shi & Herniman, 2023). The use of emojis, caps lock, and expletives reinforced the emotions in many comments while in others, the use of logical argumentation was clear.

Our findings indicate a sizable conflict on similar issues and features between people. A fundamental one is that people had conflicting assessments of Sora’s creativity and video-generation quality: while some thought Sora was a revolutionary technique capturing physical characteristics perfectly, others identified imperfections in physical simulation and noticed similarities between Sora-generated videos and Midjourney-generated pictures. These conflicts were also present in people's views on the desirability of the technology, its impact on employment, AI law enforcement, its impact on content democratization, extent of AI regulation, and trust of OpenAI.

Sora could also be viewed as a continuation of generative AI models that have captivating people's attention in the past few years. An interesting parallel can be found with Virtual Influencers, which are artificially generated entities acting as social media influencers, often using hyper-realistic AI-generated imagery, but still human-curated: many of the thoughts evoked in this study included science-fiction themes, blurring of lines between reality and fiction, democratization using AI, and competition between artificially generated content and human-generated content, resonating with quotes of long-term followers of virtual influencers (Choudhry et al., 2022). This is indicative of something more meta: the public sentiment, perception, and expectations of an individual AI technology may be influenced less by the specifics of that particular product and more by the general zeitgeist of AI. The conversations about Sora were less about the features of Sora itself but more about its potential impact on society, jobs, and freedom implicating that such fundamental debates on AI have existed in the public sphere (Calhoun, 1993) for some time, suggesting that such public opinion could potentially be harnessed in creating fair legislation as has been envisaged previously (Leane, 2010), and resonating with a more flexible legislative approach (Wiener, 2004).

## Policy Recommendations

### *AI Watermarking*

Building on results regarding legal enforcement of labeling requirements, it is necessary to clarify what the mandate is as much as it is to articulate procedures, rights, and responsibilities regarding enforcement mechanisms. One major governance recommendation from this work is labeling AI-generated content; AI watermarking is one approach to labeling by verifying authenticity, protecting intellectual property, and preventing misinformation or mislabeling (Fu et al., 2024; Regazzoni et al., 2021). Such a labeling approach places the burden of compliance on the organizations and researchers that produce generative AI models, services, and products rather than on end-users with varying degrees of technical literacy. This centralized compliance mechanism easily facilitates enforcement, disambiguates provenance, and provides clarity to users of AI systems and for downstream information seekers.

### *Reapplying Existing Laws*

A second governance recommendation from this project is to leverage existing law and apply it in new contexts, such as anti-discrimination law to address issues of bias, harassment, and equity, along with obscenity and child-protection laws to protect minors with respect to pornography. Creative applications of broadly and flexibly written laws as pertain to similar harms or governance challenges irrespective of technical system, jurisdiction or context, or regulatory domain can help to address emergent harms, as identified in this research, as well as around emerging technologies more broadly. Such efforts have already been useful in protecting victims of revenge porn and deepfakes (Levendowski, 2013; 2023).

### *Accessibility*

A third governance recommendation pertains to accessibility standards as applied to AI systems; concerns expressed about potential guardrails infringing upon speech or creativity, as well as potentially making systems unusable, can be assuaged, reconciling them with accessibility interests, given decades of research demonstrating the positive impact of accessibility standards and testing on usability for all (e.g., Petrie & Kheir, 2007).

### *An Entrepreneurial Approach to Regulation*

Literature from environmental technologies and their regulation indicates that technology innovation is reduced by regulation, but regulation has a positive effect on public health (Zhou et al., 2021). This tradeoff between public benefit and technological innovation resulting from regulation can also be seen in the views on technology regulation expressed by the two largely non-intersecting inter-disciplinary silos of law and technology, and law and economics (Butenko & Larouche, 2015). Law and economics assumes that innovation is good for welfare, while law and technology is more critical of innovation and also views it largely as exogeneous to the regulatory process. While law and technology literature wishes that law intervenes early in emerging innovations, law and economics posits the very opposite; it is thus desirable to balance these opposing regulatory perspectives (Butenko & Larouche, 2015). Technology regulation is a highly complex undertaking, and it requires an entrepreneurial approach with flexibility and empirical testing, constituting "regulatory property" to protect against market failures but guard against overprotection, fostering an ecology that allows diffusion of regulatory innovations, adapting the prevailing theory of regulatory politics, engaging of global economies, and the rewarding of policy innovation (Wiener, 2004). This entrepreneurial and nimble approach to regulation will resolve the tradeoff between "unethical AI" and "unusable AI" in the regulation of generative AI models.

## CONCLUSION

Through a thematic analysis of social media discussions about Sora and relevant generative AI tools, we gauged people's reactions to and concerns about Sora and regulatory solutions. We synthesized policy recommendations drawing on and expanding public perceptions, including using AI watermarking to verify authenticity and protect intellectual property, leveraging existing laws to combat pornography and bias, and addressing the tension between ethics and usability of AI systems.

There are several limitations of this research. First, we leveraged a qualitative social media analysis to probe people's perceptions of Sora's societal impact and ethical concerns. We do not aim to provide a generalizable account of the complicated GenAI landscape. Future research could consider a larger-scale quantitative analysis to capture the evolving sentiment and perceptions around AI hype, and people's contrasting views as evidenced by our data. Second, a key argument from our analysis is that Sora is not as creative as humans, but only replicates what is already there. Bellini-Leite (2024) proposed a thought experiment to probe the creativity of Sora, which is an important direction to delineate Sora's potential impact on creativity and creative jobs. Third, while artists expressed concerns about their job prospects in the GenAI era, future research could consider interviewing different professionals to understand their perspectives.

## GENERATIVE AI USE

We confirm that we did not use generative AI tools/services to author this submission.